
\input phyzzx
\input epsf
\FRONTPAGE
\line{\hfill BROWN-HET-1001}
\line{\hfill September 1995}
\vskip1.5truein
\titlestyle{{FORMATION OF HIGH REDSHIFT OBJECTS IN A COSMIC STRING
THEORY WITH HOT DARK MATTER}}
\bigskip
\author{R. Moessner$^{1,2,}$\footnote{a)}{email:
moessner@het.brown.edu} and R.
Brandenberger$^{1,3,}$\footnote{b)}{email: rhb@het.brown.edu}}
\centerline{1)~ {\it Brown University, Department of Physics}}
\centerline{{\it Providence, RI 02912, USA}}
\centerline{2)~ {\it Max Planck Institut f\"ur Astrophysik}}
\centerline{\it D-85740 Garching, Germany}
\centerline{3)~ {\it Physics Department, University of British
Columbia}}
\centerline{\it Vancouver, B.C. V6T 1Z1, Canada}
\bigskip
\abstract
Using a modification of the Zel'dovich approximation adapted to hot
dark matter, the accretion of such matter onto moving cosmic string
loops is studied.  It is shown that a large number of $10^{12}M_\odot$
nonlinear objects will be produced by a redshift of $z=4$.  These
objects could be the hosts of high redshift quasars.
\endpage
\chapter{Introduction}
Observation of high redshift objects has emerged as a powerful tool
for testing theories of structure formation.  For example, in
inflationary Universe models, the epoch when the first massive
nonlinear structures (which could be the hosts of quasars or
primordial galaxies) form is a sensitive function of the fraction of
hot dark matter.  Recent data on the abundance of damped Lyman alpha
absorption systems (DLAS)$^{1)}$ and on the quasar abundance$^{2)}$
has provided tight constraints on inflation-based models of structure
formation with adiabatic density fluctuations and mixed dark matter.

Searches for high redshift quasars have been going on for some
time$^{2)}$.  The quasar luminosity function is observed to rise
sharply as a function of redshift $z$ until $z \simeq 2.5$.
According to recent results from the Palomar grism survey by Schmidt et al.
(1995)$^{2)}$, it peaks in the redshift interval $z \, \epsilon$ [1.7, 2.7]
and declines at higher redshifts. Irwin et al.$^{2)}$ on the other hand
find that the luminosity function is constant up to redshifts of about $4$.
Quasars (QSO) are extremely luminous, and it is generally assumed that they are
powered by accretion onto black holes. It is possible to estimate the mass
of the host galaxy of the quasar as a function of its luminosity,
assuming that the quasar luminosity
corresponds to the Eddington luminosity of the black hole. For a quasar
of absolute blue magnitude $M_B= -26$, the host galaxy mass can be
estimated as$^{3)}$
$$
M_G=c_1 10^{12} M_\odot \, ,  \eqno\eq
$$
where $c_1$ is a constant which contains the uncertainties in relating
blue magnitude to bolometric magnitude of quasars, in the baryon
fraction of the Universe and in the fraction of baryons in the host
galaxy able to form the compact central object (taken to be 10$^{-2}$).
The best estimate for $c_1$ is about 1.
Models of structure formation have to pass the test of producing
enough early objects of sufficiently large mass to host the observed
quasars.

Damped Lyman alpha systems provide potentially even more powerful
constraints on structure formation models.  Evidence is
mounting$^{4)}$ that the absorption line systems observed in the
spectrum of distant quasars are due to progenitors of typical
galaxies.  Based on the number density of absorption lines per
frequency interval and on the column density calculated from
individual absorption lines, the fraction of $\Omega$ in bound neutral gas
(denoted by $\Omega_g$) can be estimated.  Recent observational results$^{1)}$
indicate that
$$
\Omega_g (z) > 10^{-3} \eqno\eq
$$
for $z \, \epsilon$ [1,3], with the highest value in fact taken on at
$z = 3$ ! Note that the above corresponds to a value $\Omega$ in bound
matter which is larger by a factor of $f^{-1}_b$, where $f_b$ is the
fraction of bound matter which is baryonic.  In a $\Omega =1$
cosmology, the value of $f_b$ is expected to be of the order $10^{-1}$.

The constraints coming from observation of high redshift objects for
inflation-based models of structure formation were studied by several
groups.  In Ref. 7, it was shown that the high redshift quasar
abundance is compatible with an unbiased cold dark matter (CDM)
model\footnote{c)}{A model with flat spectrum $(n=1), \, \Omega = 1$,
and vanishing cosmological constant $\Lambda$.}, but that the theory predicts
an exponential decrease of
the quasar abundance for $z > 5$.  Recently$^{8)}$, this analysis has
been extended to mixed dark matter (MDM) models.  The abundance of
damped Lyman alpha systems was used in Refs. 3, 5 \& 6 to further
constrain MDM models.  It was found that models with a fraction of hot
dark matter (HDM) exceeding $\Omega_{\nu} = 0.2$ do not predict enough
nonlinear structures at high redshift in order to be able to explain
the data.

However, there exists a viable class of alternative theories of
structure formation, the cosmic string (CS) models.  In these models,
density fluctuations are generated by topologically stable strings of
trapped energy density (one-dimensional topological defects in a
relativistic field theory describing matter), instead of originating
as the result of quantum fluctuations during an early period of
exponential expansion of the Universe.  The main purpose of this paper
is to present a preliminary study of the constraints on cosmic string
models which can be derived using the QSO and DLAS abundances.
The model we are most interested in is a cosmic string
theory in a spatially flat $(\Omega =1)$ Universe with only HDM and
baryons.  This model is briefly reviewed in Section 2.

Our main result is that for reasonable values of the parameters of the
model, the CS and HDM theory is compatible with the present
observational constraints on the quasar abundance of the Palomar
grism survey by Schmidt et al.$^{2)}$, which are plotted in
Figure~1. We will compare these observations for the comoving
space density of quasars brighter than $M_B= -26$ with the abundance of
objects of mass greater than $M_G$ (given in (1.1)) in
the cosmic string and hot dark matter model..
We also comment on the implications of the DLAS abundance (as given by
(1.2)) for cosmic strings.

Our result is quite nontrivial.  The reason why inflationary MDM
models with a large fraction of HDM do not predict a sufficient
abundance of nonlinear objects at high redshifts is that the spectrum
of density perturbations is suppressed at small wavelengths (those
which first become nonlinear in the standard CDM inflationary model)
by neutrino free streaming$^{9)}$.  The primordial power spectrum in a
cosmic string theory is scale invariant$^{10)}$, as it is in
inflation-based models.  The reason why a CS and HDM model is viable
at all is that cosmic strings survive free
streaming$^{11)}$.  Since the strings are long-lived seed
perturbations (as opposed to adiabatic dark matter fluctuations),
accretion of dark matter on small scales -- wavelength $\lambda <
\lambda^{max}_J$, where $\lambda^{max}_J$ is the maximal
neutrino free streaming scale (whose value is given later) -- is
delayed but not prevented.  Thus, the spectrum of density
perturbations is not exponentially cut off below $\lambda^{max}_J$
as it is for an inflationary HDM theory.  It is$^{12)}$, however,
suppressed by a power of $\lambda/\lambda^{max}_J$ compared to that
of an inflationary CDM model.  It also has a smaller amplitude than
the MDM model with $\Omega_{\nu} = 0.2$.  Hence, it would seem that the CS
and HDM model would be unable to explain the abundance of high
redshift $QS0$ and DLAS. However, as explained in the following
paragraph, the above reasoning misses a crucial point.

In the cosmic string theory -- in contrast to inflation-based
models -- the density field is not a Gaussian random field.  There are
localized high density peaks even when the average density contrast is
small.  Hence, knowledge of the density power spectrum is insufficient
to calculate the number density of nonlinear objects.

In particular, cosmic string loops seed large amplitude local density
contrasts.  In this paper, we study the accretion of hot dark matter
onto moving string loops and use the results to compute the number
density of high redshift objects as a function of a parameter $\nu$
which determines the number density of loops in the scaling solution
(see Section 2).  We demonstrate that for realistic values of $\nu$, the
number of massive nonlinear objects at redshifts $\le 4$ satisfies
the recent observational constraint of quasar abundances (see Figure~1).  We
also comment on the implications of (1.2).

The next section of this paper contains a brief review of the CS and
HDM theory of structure formation.  In Section 3 we summarize the
methods used: a Zel'dovich approximation$^{13)}$ technique
modified$^{14)}$ to HDM and its adaptation to moving seed
perturbations$^{15)}$.  Section 4 contains the main calculations, and
in Section 5 we discuss the results. Units in which $\hbar = c= k_B =
1$ are used throughout, and a Hubble constant of $H=50 \,h_{50} {\rm km s^{-1}
Mpc^{-1}}$ and a redshift at equal matter and radiation of
$z_{eq}=5750 \,\Omega h_{50}^{2}$ are used.

\chapter{Brief Review of the Cosmic String and Hot Dark Matter Theory}

Cosmic strings$^{16)}$ are one-dimensional topological defects which
are predicted in many relativistic field theories describing matter.
In such theories, a network of strings forms during a phase transition
in the very early Universe.  These strings are characterized by a mass
per unit length $\mu$ (which in principle is the only free parameter
in a cosmic string model) which determines their gravitational
effects.

After the time of formation, the network of strings rapidly approaches
a ``scaling solution", a distribution of defects whose statistical
properties are time independent when lengths are rescaled by dividing
by the Hubble radius.  The mean separation of the strings increases by
having strings interconnect and chop off loops.  The scaling solution
of the string network is characterized by a fixed number $N$ of long
strings crossing each Hubble volume at any time $t$, and the presence of
loops with a distribution
$$
n (l,t) = \nu l^{-2} t^{-2} \, , \eqno\eq
$$
where $l$ is the length of the loop and $\nu$ is a constant.  The
quantity $n (l,t) dt$ gives the number per unit physical volume at time
$t$ of loops with lengths in the interval between $l$ and $l + dl$.
The loops are remnants of string intercommutations at times $t^\prime
< t$.  Loops oscillate and decay slowly by emitting gravitational
radiation.  Hence, there is a lower cutoff value of $l$ for the
distribution (2.1) given by
$$
l_{min} \sim G \mu t \, , \eqno\eq
$$
$G$ being Newton's constant.  Below $l_{min}$, the distribution
$n(l,t)$ becomes constant.  We shall not discuss this point since it
will not be relevant to our computations.

In principle, the properties of the scaling solution and hence also
the value of the constant $\nu$ is calculable, albeit only numerically.
In practice, however, the dynamics of the defect network is quite
complicated and the numerical resolution inadequate to solve this
problem.  Thus, we must treat $\nu$ as an undetermined constant.  There
are two more such constants which are denoted by $\alpha$ and $\beta$.
The first constant determines the mean radius $R_f$ of a string loop
at the time of formation
$$
R_f (t) = \alpha t \, , \eqno\eq
$$
the second relates the mean radius $R$ to the length $l$ of a loop:
$$
l = \beta R \, . \eqno\eq
$$
Numerical simulations$^{17)}$ indicate that $\alpha \le 10^{-2}$ and
$\beta \simeq 10$.  They also indicate that $N \sim 10$.  From these
values it follows that -- unless $\nu$ is extremely large -- most of the
mass of the string network resides in long strings (where long strings
are defined operationally as strings which are not loops with radius
smaller than the Hubble radius).

In the cosmic string theory there are two basic mechanisms which seed
structures, loops and wakes.  String loops act gravitationally almost
like point mass objects when viewed from distances larger than $R$.
Long straight strings, on the other hand, lead to planar over-dense
regions called wakes$^{18)}$.  On distance scales smaller than its
curvature radius, the local gravitational attraction of a string
vanishes.  However, space perpendicular to the string is conical with
deficit angle $8 \pi G \mu$.  Hence, a string moving with transverse
velocity $v_s$ will impart a velocity perturbation
$$
\delta v = 4 \pi G \mu v_s \gamma (v_s) \eqno\eq
$$
towards the plane behind the string, where $\gamma (v)$ is the
relativistic $\gamma$ factor.  This develops into a planar over-dense
region behind the string, the wake.

Since most of the mass in the string network is in long strings, the
wake mechanism will be responsible for the formation of most of the
present structure in the Universe.  The thickest and most numerous
wakes are those created at the time of equal matter and radiation
$t_{eq}$$^{19)}$.  The cosmic string theory hence predicts a
distinguished scale and topology of the large-scale structure in
encouraging agreement with the data from the CfA galaxy redshift
survey$^{20)}$.

The scaling for the defect network leads to a scale invariant spectrum
of density perturbations, which in turn leads to a ``scale-invariant"
$(n = 1)$ spectrum of microwave anisotropies$^{21)}$.  Normalizing the
model by the COBE results gives$^{22)}$
$$
G\mu \simeq 10^{-6} \, . \eqno\eq
$$

The accretion of hot dark matter onto cosmic string wakes was
considered in detail in Ref. 14.  It was found that the first comoving
scale to go nonlinear about a wake caused by a string at time $t_{eq}$
is
$$
q_{max} (t_{eq}) = v_{eq} z_{eq} t_{eq} \simeq v_{eq} \cdot 50 h_{50}^{-2} \,
{\rm Mpc} \eqno\eq
$$
where $v_{eq}$ is the mean hot dark matter velocity at $t_{eq}$.  In
a $\Omega = 1$ Universe, $v_{eq}$ is about 0.1.  Hence, the distance
$q_{max}$ is in good agreement with the observed thickness of the CfA
galaxy sheets$^{20)}$.

Note, however, that the first dark matter nonlinearities about wakes
form only at late times, at a redshift$^{14)}$
$$
z_{max} = {24 \pi\over 5} G \mu v_s \gamma (v_s) v_{eq}^{-1}
z_{eq} \, , \eqno\eq
$$
which for $v_s \gamma (v_s) \simeq 1$ and for the value of $G\mu$ from
(2.6) corresponds to a redshift of about 1.  Before this redshift, no
nonlinearities form as a consequence of accretion onto a single
uniform wake.

Thus, in the cosmic string and hot dark matter theory, a different
mechanism is required in order to explain the origin of high redshift
objects.  Possible mechanisms related to wakes are early structure
formation at the crossing sites of different wakes$^{23)}$,
small-scale structure of the strings giving rise to
wakes$^{24,25,26)}$, and inhomogeneities inside of wakes$^{26)}$.  In
this paper, however, we will explore a different mechanism, namely the
accretion of hot dark matter onto loops.

In earlier work,$^{11)}$ the accretion of hot dark matter onto static
cosmic string loops was studied.  It was found that in spite of free
streaming, the nonlinear structure seeded by a point mass grows from
inside out, and that the first nonlinearities form early on (accretion
onto string filaments proceeds similarly$^{25)}$).  In the context of
the ``old" cosmic string scenario (wakes unimportant), this mechanism
was used in Ref. 27 to derive the mass function of galaxies.  Since
loop accretion leads to nonlinear structures at high redshift, we will
now investigate this mechanism in detail to see whether enough high
redshift massive objects to statisfy the QSO constraints and (1.2) form.

\chapter{Modified Zel'dovich Approximation}

We will use the Zel'dovich approximation$^{13)}$ and modifications
thereof to study the accretion of hot dark matter onto moving string
loops.  The Zel'dovich approximation is a first order Lagrangian
perturbation theory technique in which the time evolution of the
comoving displacement $\psi$ of a dark matter particle from the
location of the seed perturbation is studied.

The physical distance of a dark matter particle from the center of the
cosmic string loop is
$$
h (q, t) = a (t) [q - \psi (q,t)] \, . \eqno\eq
$$
The scale factor $a(t)$ is normalized such that $a (t_0) = 1$, where
$t_0$ is the present time.  The Zel'dovich approximation is based on
combining the Newtonian equation for $h$
$$
\ddot h = - \nabla_h \Phi \eqno\eq
$$
with the Poisson equation for the Newtonian gravitational potential
$\Phi$ and linearizing in $\psi$.  For a pointlike seed mass of
magnitude $m$ located at the comoving position $\underline{q}^\prime =
0$ the resulting equation is
$$
\ddot \psi + 2 {\dot a\over a} \dot \psi + 3 {\ddot a\over a} \psi =
{Gm\over{a^3 q^2}} \, . \eqno\eq
$$

This equation describes how as a consequence of the seed mass, the
motion of the dark matter particles away from the seed (driven by the
expansion of the Universe) is gradually slowed down.  If the seed
perturbation is created at time $t_i$ and the dark matter is cold,
then the appropriate initial conditions for $\psi$ are
$$
\psi (q,t_i) = \dot \psi (q,t_i) = 0 \, , \eqno\eq
$$
leading -- for $a (t) \sim t^{2/3}$ appropriate in the matter
dominated epoch $t > t_{eq}$ -- to the solution
$$
\psi (q,t) = {9\over{10}} Gm \, \left({t_0\over q}\right)^2
\left({t\over t_i}\right)^{2/3} \, . \eqno\eq
$$

As formulated above, the Zel'dovich approximation only works for cold
dark matter, particles with negligible thermal velocities.  The theory
of interest to us, however, is based on hot dark matter.  Luckily, the
Zel'dovich approximation can be modified for HDM$^{14)}$.  HDM
particles have large thermal velocities.  At time $t$, the free
streaming length in comoving coordinates is
$$
\lambda_J (t) = v (t) z (t) t \, , \eqno\eq
$$
where $v(t) \sim z (t)$ is the hot dark matter velocity. The length
$\lambda_J (t)$ is the mean distance an HDM particle will move in
one expansion time.  Free streaming erases density perturbations on
scales $q < \lambda_J (t)$, a scale which decreases as $t^{-1/3}$
as $t$ increases.

A simple prescription\footnote{d)}{Note that this prescription has
been shown$^{14)}$ to give good agreement with an analysis obtained by
tracking the full phase space distribution of HDM particles by means
of the linearized collisionless Boltzmann equation.} for taking into
account free streaming in the Zel'dovich approximation is to -- for a
fixed comoving scale $q$ -- only let the perturbation start to develop
at time $t_s (q)$ when
$$
\lambda_J (t_s (q)) = q \, , \eqno\eq
$$
i.e., replace the initial conditions (3.4) by
$$
\psi (q, \tilde t_s (q)) = \dot \psi (q, \tilde t_s (q)) = 0 \eqno\eq
$$
with
$$
\tilde t_s (q) = {\rm max} \, \{ t_i, t_s (q) \} \, . \eqno\eq
$$
The result for the comoving displacement $\psi (q)$ then becomes
$$
\psi (q,t) = {9\over 10} Gm \left( {t_0\over q}\right)^2
\left({t\over{\tilde t_s (q)}}\right)^{2/3} \, . \eqno\eq
$$

We can now define the mass that has gone nonlinear about a seed
perturbation as the rest mass inside of the shell which is ``turning
around", i.e. for which
$$
\dot h (q, t) = 0 \, . \eqno\eq
$$
This yields an equation
$$
q = 2 \psi (q,t) \eqno\eq
$$
for the scale $q_{nl}(t)$ which is turning around at time $t$.  For
cold dark matter, Eq. (3.5) can be combined with Eq. (3.12) to obtain
$q_{nl} (t)$ as well as the corresponding mass
$$
M_{CDM} (t) = {2\over 5} m \left( {t\over t_i}\right)^{2/3} \, .
\eqno\eq
$$
Note the similarity of this result to what can be obtained in linear
perturbation theory: $(t/t_i)^{2/3}$ is precisely the growth factor of
linear cosmological perturbations on small scales.

For HDM, Eqs. (3.8), (3.10) and (3.12) can be combined to yield
$$
M_{HDM} (t) = {8 \over 125} \, {m^3\over{M_{eq}^2}} \left({t\over
t_{eq} } \right)^2 \eqno\eq
$$
with
$$
M_{eq} = {2\over 9} \, {v_{eq}^3 t_{eq}\over G} \, . \eqno\eq
$$

A further complication is due to the finite velocity of the cosmic
string loops.  This implies that we must extend the Zel'dovich
approximation technique to moving sources.  For CDM, this was done by
Bertschinger$^{15)}$ with the interesting result that there is to a
first approximation no change in the total mass accreted.  The
suppression of the growth of perturbations in the direction
perpendicular to the direction of motion of the seed mass is cancelled
by the larger length of the nonlinear region in the direction of motion.
For HDM, however, there will be a net suppression of growth if the
seed mass is moving.  It will be important for us to take this effect
into account.

Given a moving point source, the basic Zel'dovich approximation
equation (3.3) becomes a vector equation
$$
\underline{\ddot \psi} + 2 {\dot a \over a} \underline{\dot \psi} + 3
{\ddot a\over a} \underline{\psi} = {Gm(\underline{q} -
\underline{q}^\prime)\over{a^3 | \underline{q} -
\underline{q}^\prime|^3}} \eqno\eq
$$
where $\underline{q}^\prime (t)$ indicates the comoving position of the
source.  Without loss of generality we can take the source to move
along the $z$ axis with initial velocity $v_i$ at time $t_i$, so that$^{15)}$
$$
\underline{q}^\prime (t) = 3 v_i t_i \left(1 - \left({a(t)\over{a
(t_i)}}\right)^{-1/2} \right) \underline{e}_z z (t_i) \, , \eqno\eq
$$
$\underline{e}_z$ being the unit vector along the $z$ axis.  In the
matter dominated epoch Equation (3.16) can be solved exactly$^{15)}$
(taking $q_y = 0$ without loss of generality)
$$
\underline{\psi} (\underline{q}, t) = b (t) d_i [ f_x (\underline{q}, t)
\underline{e}_x + f_z (\underline{q} , t) \underline{e}_z ] \eqno\eq
$$
with
$$
\eqalign{
d_i & = 3 v_i t_i z (t_i) \cr
b (t) & = {1\over{15}} {Gm\over{v_i^3 t_i}} \, {a (t)\over{a (t_i)}} }
\eqno\eq
$$
and where $f_x$ and $f_z$ are known functions of $\underline{q}, t$ and
$d_i$ which at late times become independent of time and contain the
information about the geometry of the pattern of accretion onto the
moving loop.  In particular, the transverse displacement at late times
approaches
$$
f_x (\underline{q}) = \left( {d_i\over{q_x}}\right)^2 \left[
{q_x\over{d_i^2}} (R_\infty - R_i) + {q_x q_z\over{R_i d_i}} \right]
\equiv {1\over 2} \left( {d_i\over{q_x}} \right)^2 k (\underline{q}) \,
. \eqno\eq
$$
In the above,
$$
R (\underline{q}, t) = [q_{x}^2 + (q_z - q^\prime_z (t) )^2 ]^{1/2} \, ,
\eqno\eq
$$
$R_i=R(\underline{q},t_i)$, $R_\infty=R(\underline{q},t_\infty)$,
and $k (\underline{q})$ is the factor by which the accretion at
$\underline{q}$ is suppressed due to the motion of the source.

It is easy to check that for $q^2 \gg d^2_i$ the factor $k(\underline{q})$
tends to
1 (for $q^2_x \gg q^2_z$) and that the result for $\psi (q,t)$ from (3.18) goes
to the result of
(3.5).  For rapidly moving loops we are interested in the opposite
limit $d^2_i \gg q^2$.  In this case, evaluated at $q_z = 0$, the
suppression factor becomes
$$
k (q_x) \rightarrow 2 {q_x\over{d_i}} \, . \eqno\eq
$$
In this case, the ``turn-around" condition
$$
q_x = 2 \psi_x \eqno\eq
$$
for transverse accretion yields
$$
q^2_x \simeq {18\over 5} Gm t_i \left( {t_0\over{d_i}} \right) \,
\left( {t\over{t_i}} \right)^{2/3} \, . \eqno\eq
$$
For a cosmic string loop formed at time $t_i$ whose mass is (see (2.3) and
(2.4))
$$
m = \alpha \beta \mu t_i \eqno\eq
$$
this gives
$$
q_x (t_i , t) = \left({6\over 5} \alpha \beta \right)^{1/2}
(G \mu)^{1/2} v_i^{-1/2} t_0 z^{-1/2}(t)   \, . \eqno\eq
$$
{}From this result we can immediately recover Bertschinger's
result$^{15)}$ that accretion of CDM onto a moving loop is not
suppressed: the total accreted mass $M(t)$ is proportional to
$$
M(t) \sim q_x^2 (t) d_i \rho_c \eqno\eq
$$
where $\rho_c$ is the background comoving energy density.  The factors
of $v_i$ evidently cancel!

We are interested in the accretion of HDM onto moving loops.  Provided
that the turn-around distance $q_x$ of (3.26) exceeds the initial
comoving free streaming distance, i.e.
$$
q_x (t_i, t) > \lambda_J (t_i) \eqno\eq
$$
then the above analysis can also be applied to HDM.  We will check
this condition in our calculations in the following section.

\chapter{Computations}

At this point, we are able to use the methods described in the
previous section to compute the number density $n_G(>M_1,t)$ in
nonlinear objects heavier than $M_1$, and the fraction $\Omega_{nl}$ of
the critical density in such objects at high redshift $z(t)$ in the cosmic
string and hot dark matter model.  By considering only the accretion
onto string loops we will be underestimating these quantities.

It can be shown$^{27)}$ that in an HDM model string loops accrete
matter independently, at least before the large-scale structure turns
nonlinear at the redshift given by (2.8) which is about 1, i.e., later
than the times of interest in this paper.  Hence, the number density
$n(l,t)$ of loops of length $l$ given by (2.1) can be combined with the
mass $M(l)$ accreted by an individual loop (which follows from (3.13)
and (3.14)) to give the mass function $n(M,t)$.  Here, $n(M,t) dM$ is
the number density of objects with mass in the interval between $M$
and $M+dM$ at time $t$.  This in turn determines the comoving density in
objects of mass $M>M_1$,
$$
n_G(>M_1,t)=z^{-3}(t)  \int\limits^{M_2}_{M_1} dM n (M)  \eqno\eq
$$
and the fraction of the critical density in objects of mass greater
than $M_1$, $\Omega_{nl} (t) \, :$
$$
\Omega_{nl} (t) = 6 \pi G t^2 \int\limits^{M_2}_{M_1} dM Mn (M) \, . \eqno\eq
$$
If we want to compare with the observational results from QSO counts we must
use appropriate integration limits $M_1$
and $M_2$ in (4.1) and (4.2).  The mass $M_1$ is the mass limit
corresponding to the limiting QSO luminosity of the observational
sample given in (1.1), $M_1 = c_1 10^{12} M_\odot$.
For comparison with damped Lyman alpha systems, the lower cut-off mass $M_1$
is somewhat smaller, but we will not need the exact value.
For large loop radii, the approximation of treating the loop as a
point mass breaks down.  This will lead to a time dependent upper mass
cutoff $M_2 (t)$.  A rough criterion for $M_2 (t)$ can be obtained by
demanding that the mass $M(t)$ accreted onto a loop exceed the mass
in a sphere of radius equal to the loop radius at $t_i$,
$$
M (t) > {1\over{6 \pi G t_i^2}} \, {4 \pi\over 3} \, (\alpha t_i)^3 \,
. \eqno\eq
$$
Using the CDM mass formula (3.13) with initial mass (3.25) yields the
estimate
$$
M_2 (t) \simeq {2\over 5} \left( {9\over 5}\right)^{1/2} (\beta G
\mu)^{3/2} {t_0\over G} z (t)^{-3/2} \, . \eqno\eq
$$
Since
$$
{t_0\over G} \simeq 8 \cdot 10^{22} h_{50}^{-1} M_\odot \, , \eqno\eq
$$
the value of $M_2$ is much greater than both $M_1$ and the maximal
neutrino Jeans mass $M^\prime (t)$, the largest mass which is affected
by free streaming.  This justifies the use of the CDM mass formula
(3.13).

As indicated above, there is another mass which is crucial for the
computation of $n_G(t)$ and $\Omega_{nl} (t)$, namely $M^\prime (t)$.  The
easiest way
to determine $M^\prime (t)$ is to ask for the value of $M$ for which
the mass formulas (3.13) and (3.14) intersect:
$$
{2\over 5} m \left( {t\over{t_i}}\right)^{2/3} = \left({2\over
5}\right)^3 {m^3\over{M^2_{eq}}} \left({t\over{t_{eq}}}\right)^2 \, .
\eqno\eq
$$
Using the expression (3.25) for $m$ and taking account of (3.15), one
finds
$$
M^\prime (t) = {2\over 5} \left( {5\over 9}\right)^{1/4} (\alpha \beta
G \mu)^{3/4} v_{eq}^{3/4} z (t)^{-3/4} z_{eq}^{-3/4} \, {t_0\over G}
\, , \eqno\eq
$$
from which it indeed follows that -- at least for the redshifts we are
interested in -- $M^\prime (t)$ is smaller than $M_2 (t)$.

As a self-consistency check we note that $M_1$ is larger than the mass
accreted onto a loop created at $t_{eq}$.  Hence, it is consistent to
restrict our attention to the matter epoch $t > t_{eq}$.

Since the functional form of $M(m)$ and hence $M(l)$ changes at
$M=M^\prime$, the functional forms of the mass function $n(M)$ will be
different above and below $M^\prime$.  Thus
$$
n_G(>M_1,t) =  z^{-3}(t)\left[ \int\limits^{M^\prime (t)}_{M_1} d M n_H (M)  +
\int\limits^{M_2 (t)}_{M^\prime (t)} d M n_C (M)  \right]  \, . \eqno\eq
$$
Similarly for $\Omega_{nl}(t)$,
$$
\Omega_{nl} (t) =  \left[ \int\limits^{M^\prime (t)}_{M_1} d M n_H (M) M +
\int\limits^{M_2 (t)}_{M^\prime (t)} d M n_C (M) M \right] 6 \pi G t^2
\eqno\eq
$$
where $n_H (M)$ and $n_C (M)$ respectively refer to the mass functions
computed with the HDM and CDM mass formulas (3.13) and (3.14)
respectively.  Combining (2.1) with (3.13) and (3.14) we obtain
$$
n_C (M,t) = \nu {24\over{125}} (\alpha \beta)^2 {\mu^3\over M^4} \eqno\eq
$$
and
$$
n_H (M,t) = \nu {2\over{15}} \, {z_{eq}\over z} \, {\mu\over{t^2}} \,
{1\over{M_{eq}^{2/3}}} \, {1\over{M^{4/3}}} \, . \eqno\eq
$$
As a consistency check, we note that for $M = M^\prime (t)$ the above
two expressions coincide.

By inspection of (4.8-4.11) it is clear that the integrals for
$n_G(t)$ and $\Omega_{nl}(t)$ are dominated at $M = M^\prime (t)$ and that a
reasonable approximation to (4.8) will be
$$
n_G(>M_1,t) \sim n_C (M^\prime , t) M^{\prime} z^{-3}(t) \simeq
{1\over 5} \nu (\alpha \beta)^2 {\mu^3\over{M^\prime (t)^3}} z^{-3}(t) \, ,
\eqno\eq
$$
which gives
$$
\eqalign{
n_G(>M_1,t) & \sim  5 \nu (\alpha \beta)^{-1/4} (G \mu)^{3/4}
({{z_{eq}} \over {v_{eq}}})^{9/4} z^{-3/4}(t) t_0^{-3} \cr
& \sim  2 \cdot 10^{-4} \nu  z^{-3/4}(t) h_{50}^{9/2} (h_{50}^{-1} {\rm
Mpc})^{-3} }   \eqno\eq
$$
when inserting $v_{eq} = 0.1$, $G \mu = 10^{-6}$ and the values of
$\alpha = 10^{-2}$ and $\beta=10$ from Section~2. This result is plotted in
Figure~1. Similarly, (4.9) can be approximated by
$$
\Omega_{nl} (t) \sim n_C (M^\prime , t) M^{\prime^2} 6 \pi G t^2 \simeq
{1\over 5} \nu (\alpha \beta)^2 {\mu^3\over{M^\prime (t)^2}} 6 \pi G t^2
\, , \eqno\eq
$$
which gives
$$
\eqalign{
\Omega_{nl} (t) & \sim {15 \pi\over 2} \left({9\over 5}\right)^{1/2} \nu
(\alpha
\beta)^{1/2} \left({z_{eq} G \mu\over{v_{eq}}} \right)^{3/2} z (t)^{-
3/2} \cr
& \sim 10^{-1} h^3_{50} \nu z (t)^{-3/2} } \eqno\eq
$$
for the same values of the parameters as above.

At this point it looks like the cosmic string and hot dark matter
model will produce too many quasar host galaxies at redshifts $z \sim
5$.  However, so far loop velocities have been neglected.

Loop velocities can be taken into account by incorporating the
condition (3.28).  Loops which do not satisfy this criterion will not be
able to accrete much mass.  Note that as $t_i$ increases, $q_x (t_i)$
remains constant whereas $\lambda_J (t_i)$ decreases.  Hence, the
condition (3.28) corresponds to a low mass cutoff $M_c$ in the integrals
(4.1) and (4.2).  Using (3.6) and (3.26) it follows that the inequality (3.28)
becomes
$$
t_i z(t_i) v (t_i) < \left( {6\over 5} \alpha \beta \right)^{1/2} (G
\mu)^{1/2} v_i^{-1/2} t_0 z (t)^{-1/2} \, , \eqno\eq
$$
where $v (t_i)$ is the hot dark matter velocity at time $t_i$ and
$v_i$ is the loop velocity at the same time.  After some algebra, (4.16)
translates to
$$
z (t_i) < {6\over 5} \alpha \beta G \mu z_{eq}^2 v^{-1}_i v^{-2}_{eq}
z (t)^{-1} \equiv z(t_{c})  \eqno\eq
$$
where $v_{eq}$ is the HDM velocity at $t_{eq}$.  The corresponding
cutoff mass is
$$
M_c = M (t_c) = {2\over 5} \alpha \beta G \mu z (t_c)^{-1/2} z (t)^{-1}
{t_0\over G} \, .\eqno\eq
$$
It is easy to check that $M_c$ is larger than $M^\prime$, hence
justifying the use of the CDM mass formula to obtain (4.18).  This,
however, also leads to a significant reduction in the values of
$n_G(t)$ and $\Omega_{nl}$.

The effect of loop velocities is thus to replace the estimate (4.12) by
$$
n_G(>M_1,t) \sim n_C (M_c , t) M_c z^{-3}(t) \simeq 4 \nu (\alpha \beta)^{1/2}
(G \mu)^{3/2} v_i^{-3/2} ({ z_{eq}\over{v_{eq}} })^3 z^{-3/2}(t) t_0^{-3} \,
.\eqno\eq
$$
For $\alpha = \alpha_{-2} 10^{-2}, \, \beta = 10, \, v_{eq} = 0.1$ and
$G \mu = (G \mu)_6 10^{-6}$, the result becomes
$$
n_G(>M_1,t)  \sim 4 \cdot 10^{-6} \nu z^{-3/2}(t) v_i^{-3/2} \alpha_{-2}^{1/2}
(G \mu)^{3/2}_6 h_{50}^6 (h_{50}^{-1} {\rm Mpc})^{-3}
\, , \eqno\eq
$$
which is also plotted in Figure~1 for $v_i=0.25$. Similarly,
(4.13) is replaced by
$$
\Omega_{nl} (t) \sim n_C (M_c , t) M^2_c 6 \pi G t^2 \simeq 9 \pi
\nu \alpha \beta (G \mu z_{eq} v^{-1}_{eq})^2 v^{-1}_i z^{-2} (t) \, .
\eqno\eq
$$
For the same parameter choices as above, this becomes
$$
\Omega_{nl} (t) \sim 10^{-2} \nu \alpha_{-2} (G \mu)^2_6 v^{-1}_i z (t)^{-2}
h_{50}^{4}
\, . \eqno\eq
$$
Eqs. (4.20) and (4.22) are the main results of our calculations.

As a final consistency check we must verify that $M_c (t) < M_2 (t)$.
This is indeed true provided that
$$
z (t) < {3\over 2} z_{eq} v^{-1}_{eq} v_i^{-1/2} \beta \alpha^{-1/2}
G \mu \sim 17 h^2_{50}  \eqno\eq
$$
(for $v_i = 0.25$).
However, there may be an even more restrictive condition.  Since the
accretion onto a moving string is not spherically symmetric, $M_2$
might be determined not by (4.3) but by the stronger criterion
$$
q_x (t_i , t) > \alpha t_i z (t_i) \, . \eqno\eq
$$
Using (3.26), this becomes
$$
z (t_i) > {5\over 6} \alpha \beta^{-1} (G \mu)^{-1} v_i z (t) \equiv
z_m \, . \eqno\eq
$$
In order for our results (4.20) and (4.22) to be valid, $z_m$ needs to be
smaller
than $z (t_c)$.  This will only be the case if
$$
{6\over 5} \beta G \mu z_{eq} v_i^{-1} v^{-1}_{eq} > z(t) \, , \eqno\eq
$$
which is marginally satisfied for $z=4$ if $v_i = 0.25$ and $v_{eq}=0.1$,
$$
z(t) < 3 h^2_{50} \, . \eqno\eq
$$
For values
of $z$ larger than 4, the values of $n_G(t)$ and $\Omega_{nl}$ are suppressed
beyond the
results (4.20) and (4.22) since only the tail of the loop ensemble with
velocities
smaller than the mean velocity $v_i = 0.25$ manage to accrete a
substantial amount of mass.

\chapter{Discussion}

We have studied the accretion of hot dark matter onto moving cosmic
string loops and made use of the results to study early structure
formation in the cosmic string plus HDM model.  Our main results are
expressed in Eqs. (4.20), (4.22) and (4.26).

The loop accretion mechanism is able to generate nonlinear objects
which could serve as the hosts of high redshift quasars much earlier
than the time cosmic string wakes start turning around (which for $G \mu =
10^{-6}$ and $v_s = 1/2$ occurs at a redshift of about 1).  However,
there is an upper cutoff to the redshift of large mass objects which
form by this mechanism given by Eq. (4.26) and for $v_i = 0.25$ it
corresponds to a redshift of about 4.  For larger redshifts, only the
loops with velocities sufficiently small compared to the mean loop
velocity will be able to seed nonlinear objects.  Note that this
redshift cutoff is independent of the parameters $\alpha$ and $\nu$ of
the cosmic string scaling distribution which must be obtained from
numerical simulations and are still quite uncertain.

The fraction $\Omega_{nl} (z)$ of the total mass accreted into nonlinear
objects by string loops unfortunately depends very sensitively on
$\alpha$ and $\nu$.  On the other hand, this is not surprising since the
power of the loop accretion mechanism depends on the number and
initial sizes of the loops, and the scaling relation $\Omega_{nl} \sim \nu
\alpha$ is what should be expected from physical considerations.

\epsfxsize=6in \epsfbox{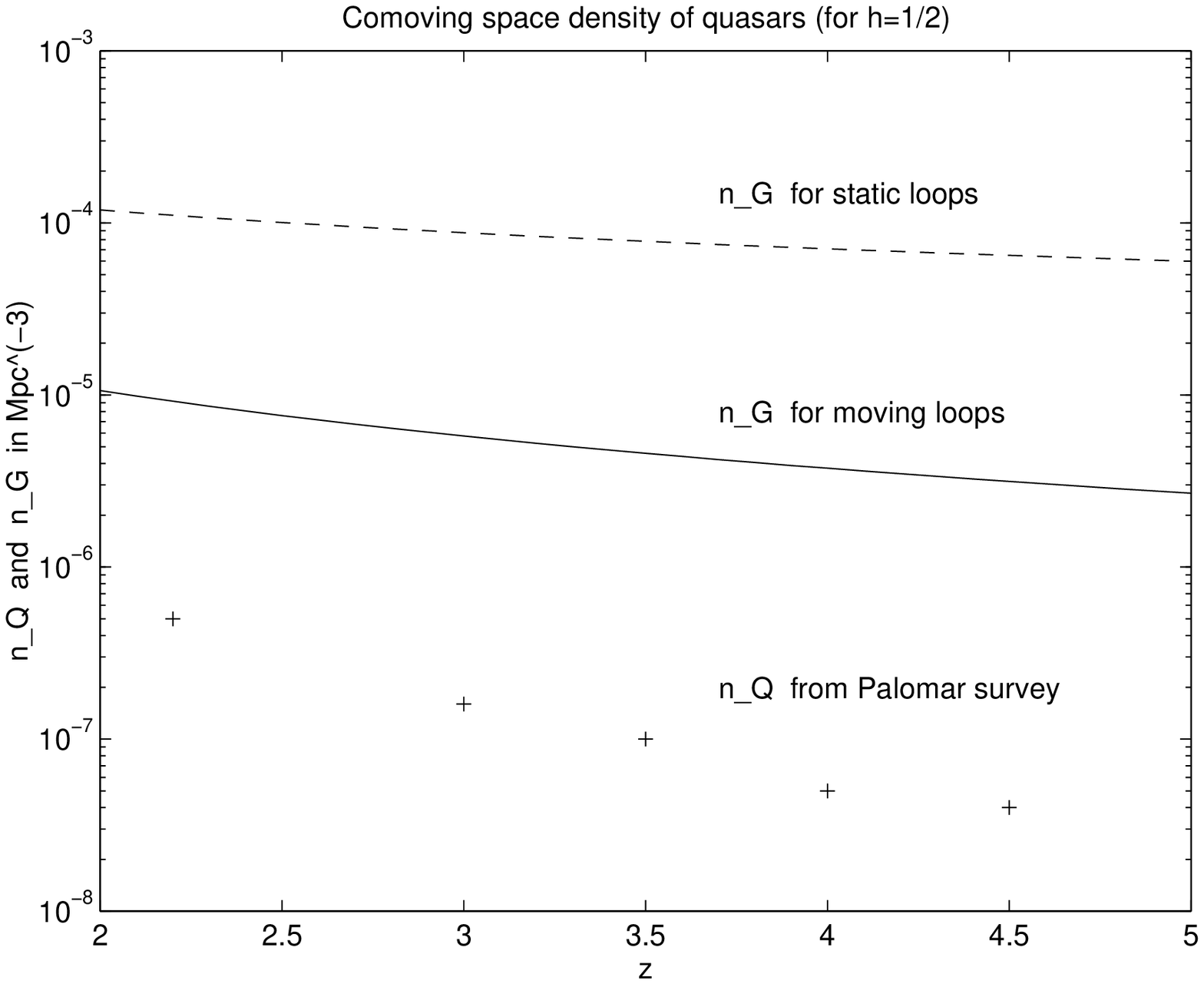}
{\baselineskip=13pt
\noindent{\bf Figure 1:} Comparison of the observed number density of quasars
(+ marks) with the number density of protogalaxies predicted in the cosmic
string theory with hot dark matter for the parameters discussed in the text.
The horizontal axis is the redshift.}

For the values $\nu =1$ and $\alpha_{-2} = 1$ which are indicated by recent
cosmic string evolution simulations$^{17)}$, we conclude from Eq. (4.20) that
the loop accretion mechanism produces enough large mass protogalaxies
to explain the observed abundance of $z \leq 4$ quasars (see Figure~1). Note
that the amplitude of the predicted protogalaxy density curves depends
sensitively on the parameters of the cosmic string scaling solution which are
still poorly determined. Hence, the important result is that there {\it are}
parameters for which the theory predicts a sufficient number of protogalaxies.
Since not all protogalaxies will actually host quasars, and since the string
parameters are still uncertain, it would be wrong to demand that the amplitude
of the $n_G$ curve agree with that of the observed $n_Q$.

It is more difficult to make definite conclusions regarding the
abundance of damped Lyman alpha absorption systems.  In the form of Eq. (1.2),
the condition for the cosmic string loop accretion mechanism to be
able to explain the data is also satisfied.  However, Eq. (1.2) refers
to the value of $\Omega$ in baryonic matter.  The corresponding
constraint on the total matter callapsed in structures associated with
damped Lyman alpha systems is
$$
\Omega_{\rm DL} (z < 3) > f^{-1}_b 10^{-3} \eqno\eq
$$
where $f_b$ is the local fraction of the mass in baryons.  From Eq.
(4.22) it follows that the above constraint is only marginally
satisfied, and this only if the local baryon fraction $f_b$ exceeds
the average value for the whole Universe of about $f_b = 0.1$.  This
could be another manifestation of the ``baryon crisis" for galaxy
clusters$^{28)}$, the fact that $f_b$ in clusters seems to exceed what
is expected based on nucleosynthesis constraints in a $\Omega = 1$
Universe.  On the other hand, in the cosmic string model with HDM we
expect $f_b$ in nonlinear objects to be enhanced over the average
$f_b$
because baryons are able to cluster during the time that the HDM is
prevented from accreting by the free streaming$^{29)}$. Thus, cosmic
strings may be able to explain the baryon excess in clusters and
restore agreement with (1.2) in a natural way.  More calculations are
required to resolve this issue.

In conclusion, we have established that in addition to being in
agreement with large-scale structure and CMB data, the cosmic string
and hot dark matter model is also able to produce a sufficient number
of protogalaxies at redshifts $z \leq 4$ which could explain the
observed abundance of quasars.  A prediction of the model is that the
distribution of these quasars should be less correlated with today's
large-scale structure than in inflation-based models, since the loops giving
rise to quasars are not correlated with the long strings present at the time of
equal matter and radiation, which give the dominant contribution to today's
large-scale structure. However, some correlation might still be present since
the loops are correlated with the long strings from which they were split off,
and since the quasar host galaxies evolve into present-day galaxies and may
fall into the potential wells created by the string wakes.
\ack
We are grateful to Houjun Mo, Joel Primack, Bill Unruh, Alex Vilenkin and Simon
White for useful
suggestions.  R. B. wishes to thank Bill Unruh and the
Physics Department of the University of British Columbia for
hospitality during the spring of 1995.  This work is supported in part by
the US Department of Energy under contract DE-FG0291ER40688, Task A
(Brown) and by the Canadian NSERC under Grant 580441 (UBC).

\bigskip
\bigskip
\REF\one{K. Lanzetta et al., {\it Ap. J. (Suppl.)} {\bf 77}, 1 (1991);
\nextline
K. Lanzetta, D. Turnshek and A. Wolfe, {\it Ap. J. (Suppl.)} {\bf 84},
1 (1993); \nextline
K. Lanzetta, A. Wolfe and D. Turnshek, {\it Ap. J.} {\bf 440}, 435
(1995);\nextline
L. Storrie-Lombardi, R. McMahon, M. Irwin and C. Hazard, `High redshift Lyman
limit and damped Lyman alpha absorbers', astro-ph/9503089, to appear in `ESO
Workshop on QSO Absorption Lines' (1995).}
\REF\two{S. Warren, P. Hewett and P. Osmer, {\it Ap. J. (Suppl.)} {\bf
76}, 23 (1991); \nextline
M. Irwin, J. McMahon and S. Hazard, in `The space distribution of
quasars', ed. D. Crampton (ASP, San Francisco, 1991), p. 117;
\nextline
M. Schmidt, D. Schneider and J. Gunn, ibid., p. 109; \nextline
B. Boyle et al., ibid., p. 191; \nextline
M. Schmidt, D. Schneider and J. Gunn, {\it Astron. J.} {\bf 110}, 68 (1995).}
\REF\three{K. Subramanian and T. Padmanabhan, `Constraints on the
models for structure formation from the abundance of damped Lyman
alpha systems', IUCAA preprint, astro-ph/9402006 (1994).}
\REF\four{F. Briggs and A. Wolfe, {\it Ap. J.} {\bf 268}, 76 (1983);
\nextline
K. Lanzetta and D. Bower, {\it Ap. J.} {\bf 371}, 48 (1992).}
\REF\five{H. Mo and J. Miralda-Escud\'e, {\it Ap. J. (Lett.)} {\bf
430}, L25 (1994).}
\REF\six{G. Kauffmann and S. Charlot, {\it Ap. J. (Lett.)} {\bf 430},
L97 (1994); \nextline
C.-P. Ma and E. Bertschinger, {\it Ap. J. Lett.} {\bf 434}, L5 (1994);
\nextline
A. Klypin et al., {\it Ap. J.} {\bf 444}, 1 (1995).}
\REF\seven{G. Efstathiou and M. Rees, {\it MNRAS} {\bf 230}, 5p (1988).}
\REF\eight{M. Haehnelt, {\it MNRAS} {\bf 265}, 727 (1993).}
\REF\nine{J. Bond, G. Efstathiou and J. Silk, {\it Phys. Rev. Lett.}
{\bf 45}, 1980 (1980); \nextline
G. Bisnovatyi-Kogan and I. Novikov, {\it Astron. Zh.} {\bf 57}, 899
(1980).}
\REF\ten{Ya.B. Zel'dovich, {\it MNRAS} {\bf 192}, 663 (1980);
\nextline
A. Vilenkin, {\it Phys. Rev. Lett.} {\bf 46}, 1169 (1981).}
\REF\eleven{R. Brandenberger, N. Kaiser, D. Schramm and N. Turok, {\it
Phys. Rev. Lett.} {\bf 59}, 2371 (1987); \nextline
R. Brandenberger, N. Kaiser and N. Turok, {\it Phys. Rev.} {\bf D36},
2242 (1987).}
\REF\twelve{A. Albrecht and A. Stebbins, {\it Phys. Rev. Lett.} {\bf
69}, 2615 (1992).}
\REF\thirteen{Ya.B. Zel'dovich, {\it Astron. Astrophys.} {\bf 5}, 84
(1970).}
\REF\fourteen{L. Perivolaropoulos, R. Brandenberger and A. Stebbins,
{\it Phys. Rev.} {\bf D41}, 1764 (1990); \nextline
R. Brandenberger, L. Perivolaropoulos and A. Stebbins, {\it Int. J. Mod. Phys.}
{\bf A5}, 1633 (1990).}

\REF\fifteen{E. Bertschinger, {\it Ap. J.} {\bf 316}, 489 (1987).}
\REF\sixteen{A. Vilenkin and E.P.S. Shellard, {\it 'Cosmic strings and other
topological defects'} (Cambridge Univ. Press, Cambridge, 1994); \nextline
M. Hindmarsh and T. Kibble, {\it Cosmic Strings'}, {\it Rep. Prog. Phys.} in
press (1995).; \nextline
R. Brandenberger, {\it Int. J. Mod. Phys.} {\bf A9}, 2117 (1994).}
\REF\seventeen{D. Bennett and F. Bouchet, {\it Phys. Rev. Lett.} {\bf 60}, 257
(1988); \nextline
B. Allen and E. P. S. Shellard, {\it Phys. Rev. Lett} {\bf 64}, 119 (1990);
\nextline
A. Albrecht and N. Turok, {\it Phys. Rev.} {\bf D40}, 973 (1989).}
\REF\eighteen{J. Silk and A. Vilenkin, {\it Phys. Rev. Lett } {\bf 53}, 1700
(1984).}
\REF\nineteen{T. Vachaspati, {\it Phys. Rev. Lett.} {\bf 57}, 1655 (1986);
\nextline
A. Stebbins et al., {\it Ap. J.} {\bf 322}, 1 (1987).}
\REF\twenty{V. de Lapparent, M. Geller and J. Huchra, {\it Ap. J. (Lett.)} {\bf
302}, L1 (1986); \nextline
M. Geller and J. Huchra, {\it Science} {\bf 246},  (1989); \nextline
V. de Lapparent, M. Geller and J. Huchra, {\it Ap. J.} {\bf 369}, 273 (1991);
\nextline
M. Vogeley, C. Park, M. Geller, J. Huchra and R. Gott, {\it Ap. J.} {\bf 420},
525 (1994).}
\REF\twentyone{N. Kaiser and A. Stebbins, {\it Nature} {\bf 310}, 391 (1984);
\nextline
R. Brandenberger and N. Turok, {\it Phys. Rev.} {\bf D33}, 2182 (1986).}
\REF\twentytwo{D. Bennett, A. Stebbins and F. Bouchet, {\it Ap. J. (Lett.)}
{\bf 399},  L5 (1992); \nextline
L. Perivolaropoulos, {\it Phys. Lett.} {\bf B298}, 305 (1993).}
\REF\twentythree{T. Hara and S. Miyoshi, {\it Prog. Theor. Phys.} {\bf 81},
1187 (1989); \nextline
T. Hara, S. Morioka and S. Miyoshi, {\it Prog. Theor. Phys.} {\bf 84}, 867
(1990); \nextline
T. Hara et al., {\it Ap. J.} {\bf 428}, 51 (1994).}
\REF\twentyfour{D. Vollick, {\it Phys. Rev. D} {\bf 45}, 1884 (1992); \nextline
T. Vachaspati and A. Vilenkin, {\it Phys. Rev Lett.} {\bf 67}, 1057 (1991).}
\REF\twentyfive{A. Aguirre and R. Brandenberger, {\it `Accretion of hot dark
matter onto slowly moving cosmic strings'}, Brown Univ. preprint BROWN-HET-995,
astro-ph/9505031 (1995),  {\it Int. J. Mod. Phys.} {\bf D}, (in press).}
\REF\twentysix{A. Sornborger, R. Brandenberger, B. Fryxell and K. Olson, `The
structure of cosmic string wakes', Brown Univ. preprint BROWN-HET-1021 (1995).}
\REF\twentyseven{R. Brandenberger and E.P.S. Shellard, {\it Phys. Rev.} {\bf
D40}, 2542 (1989).}
\REF\twentyeight{S. White et al., {\it Nature} {\bf 366}, 429 (1993); \nextline
G. Steigman \& J. Felten, {\it 'The X-Ray Cluster Baryon Crisis`} in {\it Proc.
of the St. Petersburg Gamow Seminar} (1994), ed. A. Bykov \& R. Chevalier; {\it
Sp. Sci. Rev.}, in press (Dordrecht: Kluwer).}
\REF\twentynine{R. Moessner, Brown Univ. preprint, (in preparation, 1995).}
\refout
\end